\documentclass[fp,twocolumn]{jpsj3}
\usepackage{comment}
\usepackage{txfonts}
\usepackage[dvipdfmx]{graphicx}
\usepackage{wrapfig} 
\usepackage{bm}
\usepackage{xcolor}
\makeatletter
\newcommand\newblock{\hskip .11em\@plus.33em\@minus.07em}
\makeatother

\title{Topological photonics of generalized and nonlinear eigenvalue equations}

\author{Takuma Isobe$^1$\thanks{
isobe@rhodia.ph.tsukuba.ac.jp}\thanks{Present address: {\it Technology Development Headquarters, KONICA MINOLTA, Hachioji, Tokyo 192-8505, Japan.}}, 
Tsuneya Yoshida$^{2,3}$, and Yasuhiro Hatsugai$^4$}
\inst{
$^1$Graduate School of Pure and Applied Sciences, University of Tsukuba, Tsukuba, Ibaraki 305-8571, Japan \\
$^2$Department of Physics, Kyoto University, Kyoto 606-8502, Japan\\
$^3$Institute for Theoretical Physics, ETH Zurich, 8093 Zurich, Switzerland\\
$^4$Department of Physics,
University of Tsukuba, Tsukuba, Ibaraki 305-8571, Japan
} 

\abst{
Topological photonics is developed based on the analogy of Schr\"{o}dinger equation which is mathematically reduced to a standard eigenvalue equation.
Notably, several photonic systems are beyond the standard topological band theory as they are described by generalized or nonlinear eigenvalue equations.
In this article, we review the topological band theory of this category.
In the first part, we discuss topological photonics of generalized eigenvalue equations where the band structure may take complex values even when the involved matrices are Hermitian.
These complex bands explain the characteristic dispersion relation of hyperbolic metamaterials.
In addition, our numerical analysis predicts the emergence of symmetry-protected exceptional points in a photonic crystal composed of negative index media.
In the second part, by introducing auxiliary bands, we establish the nonlinear bulk-edge correspondence under ``weak" nonlinearity of eigenvalues.
The nonlinear bulk-edge correspondence elucidates the robustness of chiral edge modes in photonic systems where the permittivity and permeability are frequency dependent. 
}

\begin{document}
\maketitle

\section{Introduction}
The topology of eigenmodes plays a central role in modern condensed matter physics.
One of the notable phenomena of topological phases is the bulk-edge correspondence where non-trivial topology results in gapless edge modes~\cite{Laughlin_IQHE_PRB(1981),Hatsugai_PRL(1993), Hatsugai_PRB(1993),Obuse_PRB(2013),Imura_PRB(2019)}
while the topological perspective is originally developed for electron systems~\cite{Klitzing_IQHE_PRL(1980),TKNN_IQHE_PRL(1982),Kohmoto_IQHE_AoP(1985),Haldane_AIQHE_PRL(1988),C.L.Kane_E.J.Mele_PRL.2005,C.L.Kane_E.J.Mele_PRL.2005_Z2,Murakami_PRL(2006),L.Fu_C.L.Kane_PRL.2007,M.Z.Hasan_C.L.Kane_RevModPhys.2010,X.L.Qi_S.C.Zhang_RevModPhys.2011,Y.Ando_JPSJ_2013,B.A.Bernevevig_T.L.Huglhes_S.C.Zhang_Science_2006,M.Knig_Science_2007,L.Fu_C.L.Kane_PRB.2007,L.Fu_C.L.Kane_PRB.2006,D.J.Thouless_PRB.1983,Schnyder_PRB.2008,X.L.Qi_T.L.Hughes_S.C.Zhang_PRB_2008,A.M.Essen_J.E.Moore_D.Vanderbilt_PRL_2009,A.Y.Kitaev_AIP_Conf_2009,S.Ryu_A.P.Schnyder_A.Furusaki_New.J.Phys_2010,S.Murakami_IOP_2007,W.Xiang_PRB_2011,Yang_PRB_2011,A.Birkov_PRL_2011,Xu_PRL_2011,Imura_PRB(2012),Kurebayashi_JPSJ_2014,N.Armitage_RevModPhys_2018,Koshino_PRB_2016},
and is applied even beyond quantum systems~\cite{Kariyado_MechGraph_Nat(2015),Yang_TopAco_PRL(2015),Huber_TopMech_Nat(2016),Susstrunk_MechClass_PNAS(2016),Tomoda_AIP(2017),Takahashi_Mech_PRB(2019),Liu_TopPhon_AFM(2020),Lee_TopCir_Nat(2018),Yoshida_Difus_Nat(2021),Makino_Difus_PRE(2022),Hu_ObsDifs_AM(2022),Delplace_TopoMeteo_science(2017),Kawaguchi_Nature(2017),Sone_Topo-Activ_PRL(2019),Knebel_GameTheor_PRL(2020),Yoshida_GameTheor_PRE(2021)}.
For instance, topological band theory is applied to photonic bands that developed topological photonics~\cite{Onoda_PRL(2004),Lu_TopPhot_Nat(2014),Khanikaev_Review_Nature(2017),Ozawa_TopPhot_RMP19, OtaIwamoto_NatPhoto(2020),Lan_Review_Elsevir(2022),Kim_3DPhC_NatComm(2022),Wang_Hybrid_NatComm(2023)}.
A photonic Chern insulator breaking time-reversal symmetry is proposed by making use of the magneto-optical effect ~\cite{Raghu_PhC_PRL(2008), Raghu_PhC_PRA(2008), MIT_PhChIns_PRL(2008),Wang_exp_Nat(2009),Ochiai_PRB(2009)} after which diverse topological bands are explored~\cite{Khanikaev_Nat(2013),Hu_TopPhot_PRL(2015),Ma_IOP(2016),Yoshimi_valleyPhC2(2020),Yoshimi_valleyPhc_Optica(2021),Lu_Nat(2013),Oono_PRB(2016),Takahashi_Optica(2017),Takahashi_JPSJ(2018),Slobozhanyuk_3d_Natphoto(2017),Obuse_PRA(2018),Yang_3d_Nature(2019),Kim_3DPhC_NatComm(2022),Hafezi_Natphoto(2011),Hafezi_NatPhoto(2013),Liang_CRtopo_PRL(2013),Mittal_PRL(2014),Rechtsman_Nature(2013),Gao_topoHMM_PRL(2015)}.

Topological band theory is further applied to systems described by a non-Hermitian matrix (e.g., dissipative systems)
, which uncovered a variety of new topological phenomena.
Such non-Hermitian systems may exhibit exceptional points (EP) protected by topology that does not have a Hermitian counterpart~\cite{Kozii_nH_arXiv(2017), Hodaei_EP_Nat(2017),Ren_EP_OptLett(2017),Shen_NHTopBand_PRL(2018),Yoshida_NHhevferm_PRB(2018),Zyuzin_NHWyle_PRB(2018),Takata_pSSH_PRL(2018),Zhou_EP_Science(2018),Ozdemir_Nat(2019),Miri_EP_Sience(2019),Michishita_EPKondo_PRB(2020),Laha_EP_ACS(2020),Mandal_HighEP_PRL(2021),Delplace_EP3_PRL(2021)}.
On an EP, the band-touching occurs for both the real and imaginary parts, which is accompanied by the coalescence of eigenvectors.
The topology of EPs is further enriched by symmetry~\cite{Zhen_ERing_nature(2015),Budich_SPERs_PRB(2019),Yoshida_SPERs_PRB(2019),Yoshida_SPERsMech_PRB(2019),Yoshida_PTEP(2020),Okugawa_SPERs_PRB(2019),Zhou_SPERs_Optica(2019),Kimura_SPES_PRB(2019)}, leading to higher-dimensional structures of EPs such as symmetry-protected exceptional rings (SPERs).
Photonic systems are one of the most potent targets of the non-Hermitian topology and attracts growing interest in terms of application, such as topological insulator laser~\cite{Harari_TILasor_Science(2018),Miguel_TILasor_Science(2018)} and EPs based high-sensitivity sensors.~\cite{Wiersig_PRA(2016),Chen_Nat(2017),Miller_PT(2017),Hodaei_Nat(2017),Zhao_Nat(2018),Chen_Nat(2018),Sakhdari_IEEE(2018),Mortensen_EP_OPTICA(2018),Dong_NatElec(2019),Zeng_OptEx(2019),Hokmabadi_Nat(2019),Lai_Nat(2019),Wiersig_EPsensor_optica(2020)}

The above significant progress of topological band theory in classical systems stems from the fact that the systems are described by the standard eigenvalue equation, providing the analogy of the Schr\"odinger equation of quantum systems.
However, notably, some of the photonic systems are beyond the standard eigenvalue problem.
Specifically, they are described by generalized eigenvalue equations (GEVEs) or nonlinear eigenvalue equations (NLEVEs) which are beyond the conventional topological band theory~\cite{Raghu_PhC_PRL(2008), Raghu_PhC_PRA(2008),IYH_PRB(2021),IYH_NanoPh(2023)}.

The aim of this article is to provide a concise review of topological photonics beyond the standard eigenvalue equations.
In the first part, we discuss topological photonics of GEVEs where the band structure may take complex values even when the matrices involved are Hermitian.
In GEVEs, those structures are caused by the indefinite property of Hermitian matrices~\cite{IYH_PRB(2021)}.
We also apply this theory to the photonic system called hyperbolic metamaterials and photonic crystals composed of negative index media.
In the second part, we establish the nonlinear bulk-edge correspondence under “weak” nonlinearity of eigenvalues by introducing auxiliary bands, which elucidates the robustness of chiral edge modes in photonic systems composed of dispersive media.

The rest of this paper is organized as follows.
In Sec.~\ref{sec:GEVE1}, non-Hermitian topological band structure caused by GEVEs is discussed.
In Sec.~\ref{sec:NLEVP0}, we discuss the topological photonics in NLEVEs.
A short summary and the remaining open questions appear at the end of this paper.

\section{Topological photonics in GEVEs}\label{sec:GEVE1}

The band structures of several photonic systems are described by GEVEs [see Sec.~\ref{sec:GEVE11}] which allows complex bands even for Hermitian matrices.
Based on topological perspective provided in Secs.~\ref{sec:GEVE12}~and \ref{sec:GEVE3},
we discuss the origin of the characteristic dispersion relations of hyperbolic metamaterials.
In addition, we discuss the emergence of SPERs in photonic crystals of negative index media which are described by GEVEs [see Sec.~\ref{sec:GEVE5}].

\subsection{Topological photonics and GEVEs}\label{sec:GEVE11}

Topological photonics explore the topological phases of electromagnetic fields. 
The behavior of electromagnetic fields is governed by Maxwell's equations, and specifically, the behavior of electromagnetic fields in photonic crystals, which are one of the representative platforms in topological photonics, is analyzed using the following equation~\cite{MIT_PhChIns_PRL(2008)},
\begin{equation}
\label{eq:maxwell eq for phc}
\sum_j\langle\phi_i|\bm{\nabla}\times\mu^{-1}(\omega,\bm{x})\bm{\nabla}\times|\phi_j\rangle\psi_j=\omega^2\sum_j\langle\phi_i|\varepsilon(\omega,\bm{x})|\phi_j\rangle\psi_j.
\end{equation}
This equation takes the form of the following generalized eigenvalue equation, 
\begin{equation}
\label{eq:GEVE1}
H\bm{\psi}=ES\bm{\psi},
\end{equation}
where $H$ and $S$ are matrices, and $E$ is the eigenvalue. 
When $S$ is the identity matrix, the problem reduces to a standard eigenvalue problem. 
In systems described by the Schr\"{o}dinger equation, the matrix $S$ represents the overlap of the basis functions, $\langle\phi_i|\phi_j\rangle$, and is always a positive definite matrix. 
However, in the context of Maxwell's equations, the matrix $S$ includes functions such as $\varepsilon$ and $\mu$, which can render $S$ an indefinite matrix.

Notably, in such cases, eigenvalues may become complex
even when the matrices are Hermitian~\cite{IYH_PRB(2021),IYH_NanoPh(2023)}.
However, previous studies in topological photonics have been primarily limited to cases where the matrix $S$ is positive definite, equivalent to the standard eigenvalue problem. 
In the following, we discuss the mechanism by which non-Hermitian topological band structures arise from indefinite Hermitian matrices in GEVEs, as well as their applications to optical systems.

\subsection{GEVEs and emergent symmetry}\label{sec:GEVE12}

Eigenvalues of Hermitian standard eigenvalue problems are limited to real.
In contrast, GEVEs can have complex eigenvalues even if the matrices involved are Hermitian~\cite{Mehl_GEVP(2004),Golub_MatComp(2013)}. 
We elucidate that the indefinite properties of both matrices $H$ and $S$ in Eq.~\eqref{eq:GEVE1} are the necessary condition of complex bands of GEVEs with Hermitian matrices. We also clarify that emergent symmetry [see Eqs.~\eqref{eq:GEVE Hsigma}~and~\eqref{eq:GEVE6}] imposes a constraint on the complex bands.

We diagonalize the matrix $S$ using a unitary matrix 
$U_{\mathrm{S}}$, 
\begin{equation}
\label{eq:GEVE2} 
H'\bm{\psi}'=ES'\bm{\psi}', 
\end{equation} 
where $H'=
U_{\mathrm{S}}^{-1}HU_{\mathrm{S}}$, $S'=U_{\mathrm{S}}^{-1}SU_{\mathrm{S}}$, and $\bm{\psi}'=U_{\mathrm{S}}^{-1}\bm{\psi}$. 
Here, $S'$ is a diagonal matrix whose diagonal components 
are denoted by $s_1, s_2, \cdots, s_n$. 
Next, we decompose the matrix $S'$ to three matrices composed of the square-root of the absolute value of $s_i$ and the sign of $s_i$,
\begin{equation}
\label{eq:GEVE3} 
S' = S^{1/2} \Sigma S^{1/2}, 
\end{equation} 
with 
\begin{align}
S^{1/2}&=
\begin{pmatrix}
\sqrt{|s_1|}& & \\ 
  &\ddots& \\
  & &\sqrt{|s_n|}
\end{pmatrix},\\ 
\Sigma&=
\begin{pmatrix}
\mathrm{sgn}(s_1)& &\\
 &\cdots& \\
 & &\mathrm{sgn}(s_n)
 \end{pmatrix}.
\end{align}
Introducing 
$\tilde{H}=S^{-1/2}H'S^{-1/2}$ and $\tilde{\bm{\psi}}=S^{1/2}\bm{\psi'}$, 
we obtain the following GEVE,
\begin{equation}
\label{eq:GEVE4}
\tilde{H}\tilde{\bm{\psi}}=E\Sigma\tilde{\bm{\psi}}.
\end{equation} 
Finally, using the relation $\Sigma^2=1$, Eq.~\eqref{eq:GEVE1} can be transformed to the standard eigenvalue equation,
\begin{equation}
\label{eq:GEVE5}
H_{\Sigma}\tilde{\bm{\psi}}=E\tilde{\bm{\psi}},
\end{equation}
with
\begin{equation}
\label{eq:GEVE Hsigma}
H_{\Sigma}=\Sigma\tilde{H},
\end{equation}
which is generically non-Hermitian.

An important observation is that the matrix $\Sigma$ is proportional to the identity matrix when the matrix $S$ is either positive or negative definite. 
In such cases, $H_{\Sigma}$ becomes a Hermitian matrix because $\Sigma$ becomes the identity matrix.
Conversely, when $S$ is indefinite, $\Sigma$ is not 
the identity matrix. 
Under these conditions, $H_{\Sigma}$ becomes non-Hermitian, resulting in complex eigenvalues. 

In a similar way, we can see that eigenvalues are real when $H$ is positive or negative definite
by rewriting Eq.~\eqref{eq:GEVE1} as 
\begin{equation}
\label{eq:GEVE inverse}
S\bm{\psi}=(1/E)H\bm{\psi}.
\end{equation}
Therefore, when matrices $H$ and $S$ are indefinite Hermitian matrices, GEVEs may have complex eigenvalues.

It should be noted that Hermiticity of $H$ results in pseudo-Hermiticity imposed on $H_{\Sigma}$
\begin{equation}
\label{eq:GEVE6}
\Sigma H_{\Sigma} \Sigma = H_{\Sigma}^\dagger.
\end{equation}
Due to the presence of pseudo-Hermiticity, the eigenvalues are given by complex conjugate pairs $(E, E^*)$ or real values $E\in \mathbb{R}$.

\subsection{Toy model analysis}
\label{sec:GEVE3}

\begin{figure}[b]
   \includegraphics[width=9cm]{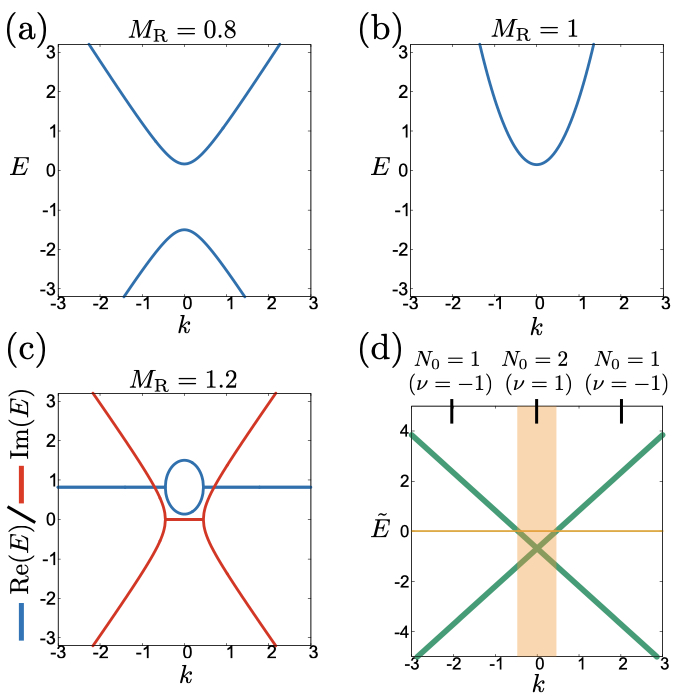}
 \caption{
(a)-(c): Band structures of Eq.~\eqref{eq:1d toy model} with $M_{\mathrm{R}}=0.8$ [(a)], $M_{\mathrm{R}}=1$ [(b)], and $M_{\mathrm{R}}=1.2$ [(c)].
(d): Eigenvalues of Eq.~\eqref{eq:matrix for zeroth chern num} and topological invariants. $N_0$ and $\nu$ represent the zeroth Chern number and the $\mathbb{Z}_2$-invariant.
}
\label{fig:1d toy model}
\end{figure}

In the above, we have discussed the conditions for complex eigenvalues of GEVEs composed of Hermitian matrices. 
Here, by analyzing toy models, we demonstrate the emergence of those complex band structures in GEVEs involving indefinite Hermitian matrices. 
Specifically, we examine the one-dimensional model described by the following equation,
\begin{equation}
\label{eq:1d toy model}
\begin{pmatrix}
M_{\mathrm{L}} & k \\
k & -M_{\mathrm{L}}
\end{pmatrix}
\bm{\psi} = E
\begin{pmatrix}
1 + M_{\mathrm{R}} & 0 \\
0 & 1 - M_{\mathrm{R}}
\end{pmatrix}
\bm{\psi},
\end{equation}
where $M_{\mathrm{L}}$, $M_{\mathrm{R}}$ are constant, and $k$ denotes the momentum (or wavenumber).
The matrix on the left-hand side becomes always indefinite when $M_{\mathrm{L}}$ is not equal to zero.
Hence, we fix $M_{\mathrm{L}}=0.3$.
The definite property of the matrix on the right-hand side depends on the magnitude of $M_{\mathrm{R}}$.
When $M_{\mathrm{R}}$ is greater than $1$, the matrix on the right-hand side becomes indefinite.

The eigenvalues of Eq.~\eqref{eq:1d toy model} are given by,
\begin{equation}
\label{eq:EV 1d toy model}
E = \frac{1}{1 - M_{\mathrm{R}}} \left[-M_{\mathrm{L}}M_{\mathrm{R}} \pm \sqrt{M_{\mathrm{L}}^2 + (1 + M_{\mathrm{R}}^2) k^2}\right].
\end{equation}
These eigenvalues are plotted in Fig.~\ref{fig:1d toy model} for each $k$.
We fix $M_{\mathrm{L}}$ to $0.3$.
When $M_{\mathrm{R}}=0.8$, eigenvalues become real, since the matrix on the right-hand side is definite.
The size of the band gap increases as $M_{\mathrm{R}}=0.8$ approaches $1$. When $M_{\mathrm{R}}=1$, one of the eigenvalues becomes infinite. This case corresponds to when the inverse matrix $S^{-1}$ cannot be defined. Notably, when $M_{\mathrm{R}}>1$, the emergence of complex eigenvalues is observed. The real (imaginary) part of the eigenvalue is plotted in blue (red). The band touching points where both the real and imaginary parts are EPs. 
These EPs are protected by the pseudo-Hermiticity discussed in the previous section.

Let us address the topological characterization of the symmetry-protected EPs by computing the zeroth Chern number~\cite{Yoshida_SPERs_PRB(2019),Yoshida_PTEP(2020)}.
In our case, the zeroth Chern number can be defined by the number of negative eigenvalues of the following Hermitian matrix,
\begin{equation}
\label{eq:matrix for zeroth chern num}
\Sigma[H_{\Sigma}(k)-E_{\mathrm{ref}}]=
\begin{pmatrix}
\frac{M_{\mathrm{L}}}{M_{\mathrm{R}}+1}&\frac{k}{\sqrt{M_{\mathrm{R}}^2-1}}\\
\frac{k}{\sqrt{M_{\mathrm{R}}^2-1}}&-\frac{M_{\mathrm{L}}}{M_{\mathrm{R}}-1}
\end{pmatrix}
-E_{\mathrm{ref}}
\begin{pmatrix}
1&0\\
0&-1
\end{pmatrix},
\end{equation}
where $E_{\mathrm{ref}}$ represents the reference point and is selected as the eigenenergy at symmetry-protected EPs. In this model, $E_{\mathrm{ref}}$ is given by
$
E_{\mathrm{ref}} = -M_{\mathrm{R}} M_{\mathrm{L}} / (1 - M_{\mathrm{R}}^2).
$
The number of negative eigenvalues changes at EPs.
In Fig.~\ref{fig:1d toy model}(d), eigenvalues of $\Sigma[H_{\Sigma}(k)-E_{\mathrm{ref}}]$ are plotted in green.
These eigenvalues become zero at the momentum where symmetry-protected EPs emerge.
The zeroth Chern number $N_0$ becomes $1$ in the white region, while it becomes $2$ in the orange region.

We note that there exists another invariant, which characterizes these symmetry-protected EPs.
The presence of the pseudo-Hermiticity allows us to define the following $\mathbb{Z}_2$-invariant~\cite{IYH_NanoPh(2023)},
\begin{equation}\label{eq:z2 invariant}
\nu =\mathrm{sgn} \Delta(\bm{k}).
\end{equation}
Here, $\Delta(\bm{k})$ is the discriminant~\cite{Yoshida_disc_PRB(2022),Wakao_disc_PRB(2022)} of the polynomial of $E$, $\mathrm{det}[H(\bm{k})-ES(\bm{k})]= \mathrm{det}[S(\bm{k})]\mathrm{det}[H_{\Sigma}(\bm{k})-E]  =a_N(-E)^N+a_{N-1}(-E)^{N-1}+\cdots +a_1(-E)+a_0$ with $a_i \in \mathbb{C}$. 
It is defined as 
\begin{equation}
  \Delta(\bm{k})=\prod_{n<n'}[E_n(\bm{k}) - E_{n'}(\bm{k})]^2,
\end{equation}
where $n$ label eigenvalues $E_n$ ($n=1,\ldots,N$).
Here, because of the pseudo-Hermiticity, $\Delta$ become real.
Furthermore, the discriminant can be computed only from the coefficients $a_i$.~\cite{footnote5} 
In Fig.~\ref{fig:1d toy model}(d), the $\mathbb{Z}_2$-invariant $\nu$ is shown in parentheses. 
It takes $1$ where $N_0=2$, while it takes $-1$ where $N_0=1$.

\begin{figure}[h]   \includegraphics[width=9cm]{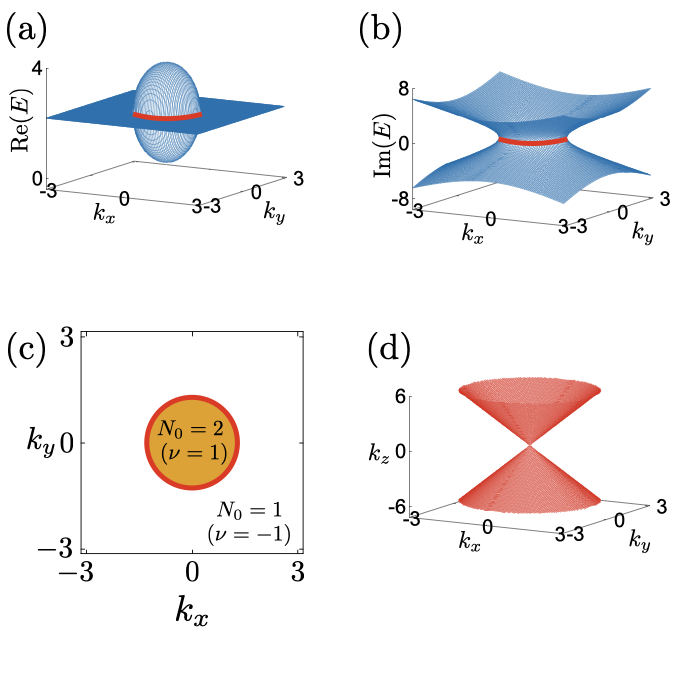}
 \caption{
(a), (b): The real and the imaginary parts
of the band structure of two-dimensional model.
 $M_{\mathrm{L}}$ and $M_{\mathrm{R}}$ are selected to $0.8$ and $1.2$.
The red lines represent the SPERs.
(c): Plot of SPERs in parameter space (red).
The zeroth Chern number ($\mathbb{Z}_2$-invariant) takes $2$ ($1$) at the orange region while it takes $1$ (-1) at the white region.
(d): Plot of the SPES in the three-dimensional model. 
$M_{\mathrm{R}}$ is selected to $0.8$. 
}
\label{fig:2d toy model}
\end{figure}

This result can be straightforwardly extended to higher-dimensional systems.
In two- (three-) dimensional systems, the symmetry-protected EPs forms lines (surface).
Here, we analyze the two-dimensional model described by the following GEVE,
\begin{equation}
\label{eq:2d toy model}
\begin{pmatrix}
M_{\mathrm{L}}&k_x-ik_y\\
kx+ik_y&-M_{\mathrm{L}}
\end{pmatrix}
\bm{\psi}=E
\begin{pmatrix}
1+M_{\mathrm{R}}&0\\
0&1-M_{\mathrm{R}}
\end{pmatrix}\bm{\psi}.
\end{equation}
This model is a
two-dimensional extension of Eq.~\eqref{eq:1d toy model}.
We fix $M_{\mathrm{L}}$ and $M_{\mathrm{R}}$ to $0.8$ and $1.2$.
The eigenvalues of this model are given by,
\begin{equation}
\label{eq:EV 2d toy model}
E=\frac{1}{1-M_{\mathrm{R}}^2}\left[-M_{\mathrm{L}}M_{\mathrm{R}}\pm\sqrt{M_{\mathrm{L}}^2+(1-M_{\mathrm{R}}^2)(k_x^2+k_y^2)}\right].
\end{equation}
In Figs.~\ref{fig:2d toy model}(a) and \ref{fig:2d toy model}(b), the real and imaginary parts of the eigenvalues are plotted.
These figures indicate the emergence of a SPER denoted by red lines.
This SPER is characterized by the zeroth Chern number [see Eq.~\eqref{eq:matrix for zeroth chern num}].
In Fig.~\ref{fig:2d toy model}(c), the zeroth Chern number is plotted.
In the orange region, the zeroth Chern number takes $2$, and takes $1$ in the white region.
The boundary between the orange region and the white region corresponds to SPER.
Therefore, the above SPER is topologically protected and robust for the perturbation.

Next, let us analyze a three-dimensional model described by,
\begin{equation}
\label{eq:3d toy model}
\begin{pmatrix}
k_z&k_x-ik_y\\
kx+ik_y&k_z
\end{pmatrix}
\bm{\psi}=E
\begin{pmatrix}
1+M_{\mathrm{R}}&0\\
0&1-M_{\mathrm{R}}
\end{pmatrix}\bm{\psi}.
\end{equation}
In this model, $M_{\mathrm{L}}$ in the one- or two-dimensional model is replaced by $k_z$.
Since the EPs emerge when the inside the square-root becomes zero, the condition for the symmetry-protected exceptional surfaces is given by,
\begin{equation}
\label{eq:SPES in 3d toy model}
k_z=\pm\sqrt{(M_{\mathrm{R}}-1)(k_x^2+k_y^2)}.
\end{equation} 
Figure~\ref{fig:2d toy model}(d) is the plot of the symmetry-protected exceptional surfaces (SPESs).
In this model, the EPs form a corn structure.

From this discussion, we can see that our results of GEVEs can be extended to two- or three-dimensional models.

\subsection{SPERs in hyperbolic metamaterials}\label{sec:GEVE4}

\begin{figure}[t]
   \includegraphics[width=9cm]{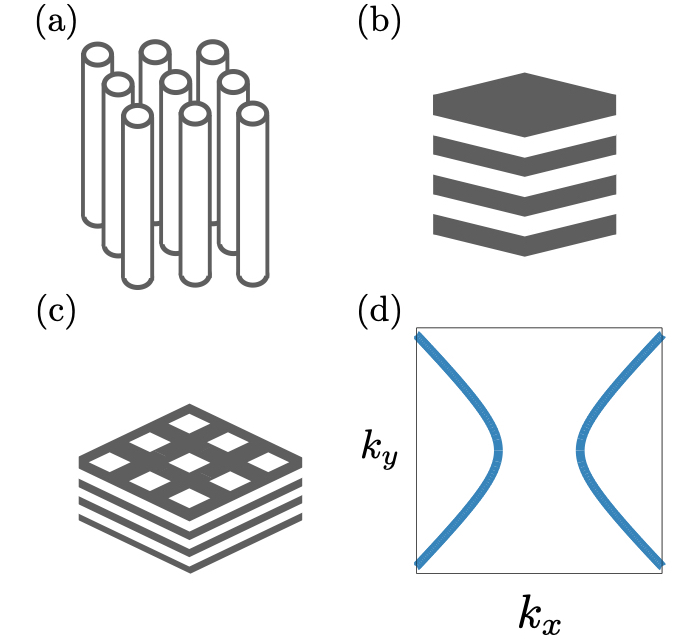}
 \caption{
	 (a)-(c): Sketch of hyperbolic metamaterials.
   (d): Isofrequency surface of hyperbolic metamaterials.
These figures are adapted with permission from Ref.~\cite{IYH_PRB(2021)}. Copyright 2024 American Physical Society.
}
 \label{fig:HMM1}
\end{figure}

In this section, we apply the above arguments to a hyperbolic metamaterial
[see Fig.~\ref{fig:HMM1}]. 
SPERs described by GEVEs provide a topological explanation for the hyperbolic dispersion observed in hyperbolic metamaterials. 

Hyperbolic metamaterials are a kind of optical metamaterial characterized by their extreme anisotropy~\cite{Smith_PRL_2003_HMMs,Smith_APL_2004_HMMs,Poddubny_NatPhoto_2013_HMMs,Drachev_OptExp_2013_HMMs,Shekhar_2014_HMMs,Ferrari_PiQE_2015_HMMs,Guo_AIP_2020_HMMs}. 
In these materials, permittivity or permeability can take on negative values in certain directions, and the isofrequency surface forms the hyperboloid. 
As a result, electromagnetic waves can pass through hyperbolic metamaterials when incident from certain directions, while they are reflected and blocked from other directions. 
The hyperbolic metamaterial can be constructed by thin wire structures~\cite{Elser_Wire-HMM_APL(2006),Mirmoosa_Wire-HMM_PRB(2016)}, metal-dielectric layered structures~\cite{Wood_Layer-HMM_PRB(2006),Caligiuri_Layer-HMM_JOP(2016)}, and fishnet structures~\cite{Kruk_NatComm_2016_HMMs}, as shown in Figure~\ref{fig:HMM1}(a)- \ref{fig:HMM1}(c).
Furthermore, recent studies have discovered natural materials corresponding to such hyperbolic systems.~\cite{Sun_Nat-HMM_AcsPhoto(2014),Narimanov_Nat-HMM_NatPhoto(2015),Zhang_Nat-HMM_Nature(2021)}

\begin{figure*}[t]
   \includegraphics[width=18cm]{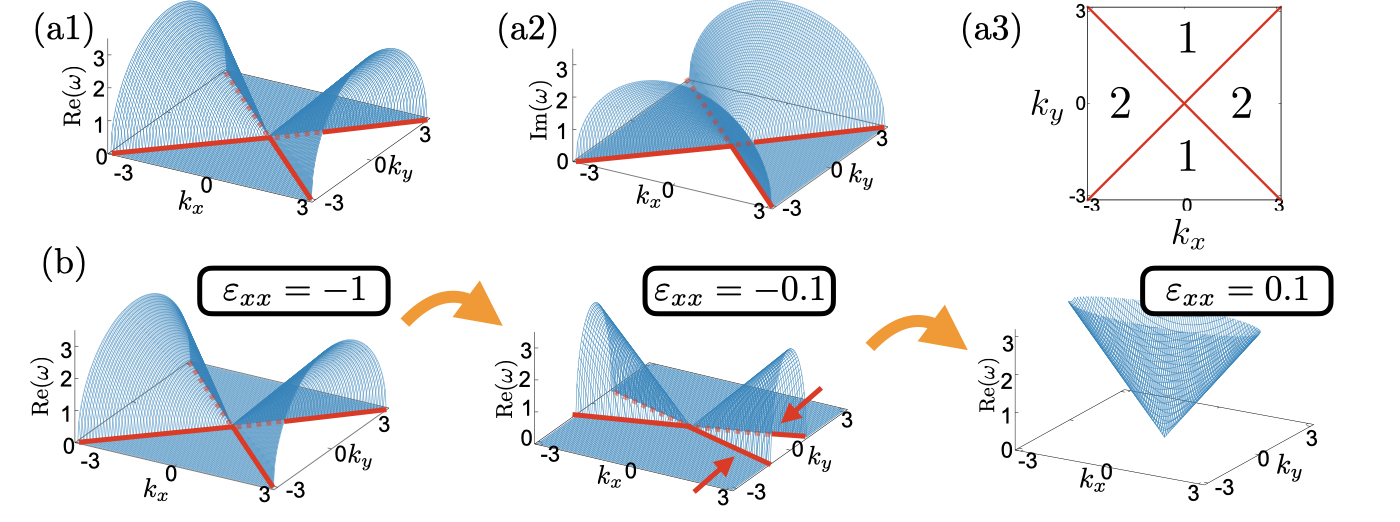}
 \caption{
	 (a1): The real part of the band structure of the hyperbolic metamaterial with $\varepsilon_{xx}=-1$, $\varepsilon_{yy}=1$, and $\mu_{zz}=1$.
   Only the positive band is plotted.
   Red lines represent SPERs.
	 (a2): the imaginary part of the band structure of the hyperbolic metamaterial with $\varepsilon_{xx}=-1$, $\varepsilon_{yy}=1$, and $\mu_{zz}=1$.
   (a3): Plot of the zeroth-Chern number.
   (b): Relation between SPERs and the magnitude of $\varepsilon_{xx}$.
   }
 \label{fig:HMM2}
\end{figure*}

We now analyze hyperbolic metamaterials and discuss the emergence of SPERs. 
Electromagnetic fields in hyperbolic metamaterials are described by the Maxwell equations. 
Here, we study a two-dimensional hyperbolic metamaterial for TE modes by analyzing the following equation,
\begin{equation}
\label{eq:3x3 HMM}
\begin{pmatrix}
0&0&-k_y\\
0&0&k_x\\
-k_y&k_x&0
\end{pmatrix}
\begin{pmatrix}
E_x\\
E_y\\
H_z
\end{pmatrix}
=\omega
\begin{pmatrix}
\varepsilon_{xx}&0&0\\
0&\varepsilon_{yy}&0\\
0&0&\mu_{zz}
\end{pmatrix}
\begin{pmatrix}
E_x\\
E_y\\
H_z
\end{pmatrix}.
\end{equation}
For simplicity, we set parameters as 
$\varepsilon_{xx}=-1$, $\varepsilon_{yy}=1$, and $\mu_{zz}=1$.
In this model, $\varepsilon_{xx}$ and $\varepsilon_{yy}$ have opposite sign and describe a hyperbolic metamaterial. 
The eigenvalues and eigenvectors are given by
\begin{equation}
\label{eq:EV HMM}
\omega_0=0, \quad\quad  \omega_{\pm}=\pm\sqrt{k_x^2-k_y^2},
\end{equation}
and
\begin{equation}
\label{eq:EVec HMM}
\bm{v}_0=\frac{1}{\sqrt{k_x^2+k_y^2}}
\begin{pmatrix}
k_x\\
k_y\\
0
\end{pmatrix}, \quad\quad
\bm{v}_{\pm}=\frac{1}{k_x}
\begin{pmatrix}
\pm\sqrt{k_x^2-k_y^2}\\
-k_y\\
1
\end{pmatrix}.
\end{equation}
Since the zero mode does not satisfy Gauss's law, we discard $\omega_0$ and $\bm{v}_0$. 
The band structure of this system is shown in Fig.~\ref{fig:HMM2} where only $\omega_+$ is plotted.
Figures~\ref{fig:HMM2}(a1) and \ref{fig:HMM2}(a2) display the real and imaginary parts, respectively. 
The isofrequency surface indeed forms a hyperboloid. 
Real and imaginary eigenvalues interchange along the red line. 
On the red lines, $\bm{v}_+$ and $\bm{v}_-$ coalesce, and $S_{3\times 3}^{-1}H_{3\times 3}$ cannot be diagonalized. 
Therefore, these red lines correspond to SPERs. 
As we have seen in the previous section, these SPERs are characterized by the zeroth Chern number. 
Figure~\ref{fig:HMM2}(a3) shows the zeroth Chern number in the parameter space. 
The zeroth Chern number changes on the SPERs, indicating that these SPERs are topologically protected and robust.

Next, let us consider increasing the magnitude of $\varepsilon_{xx}$ from negative to positive values. 
In Fig.~\ref{fig:HMM2}(b), the band structures for $\varepsilon_{xx}=-1$, $\varepsilon_{xx}=-0.1$, and $\varepsilon_{xx}=0.1$ are plotted. 
The other material parameters are fixed to 1. 
As the magnitude of $\varepsilon_{xx}$ increases, SPERs approach each other.
Despite this change, the hyperbolic dispersion persists. 
Notably, these SPERs overlap when $\varepsilon_{xx}=0$ and vanish when $\varepsilon_{xx}$ becomes positive. 
Corresponding to vanishing SPERs, the hyperbolic dispersion also vanishes simultaneously. 
This is because the hyperbolic dispersion is continuously connected to the SPERs at $\omega = 0$. 
Thus, the SPER based on GEVEs explains the origin of the hyperbolic dispersion in hyperbolic metamaterials.

\begin{figure}[b]
   \includegraphics[width=9cm]{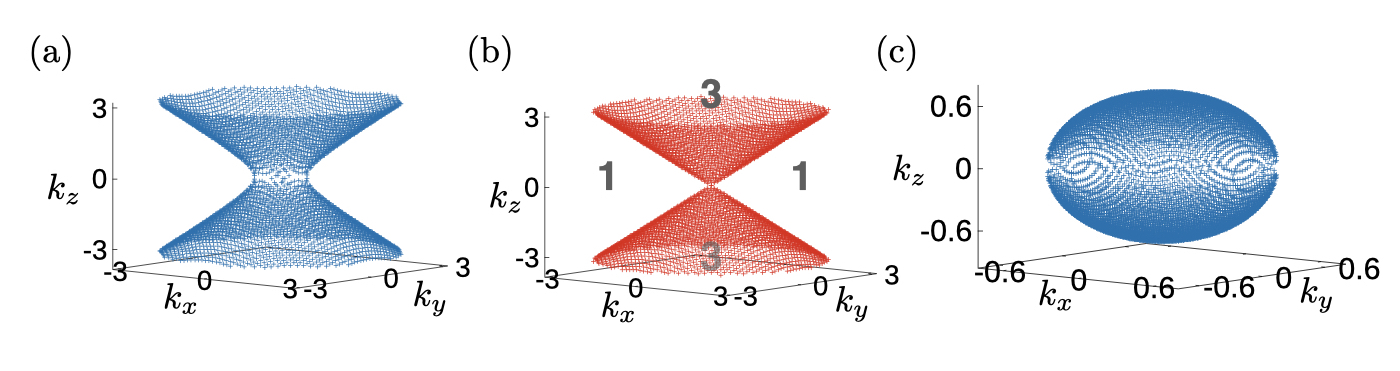}
 \caption{
 (a): The isofrequency surface of the tree-dimensional hyperbolic metamaterial with $\varepsilon_{zz}=-1$, $\varepsilon_{xx}=\varepsilon_{yy}=1$, and $\mu_{xx}=\mu_{yy}=\mu_{zz}=1$.
 (b): The SPES of the three-dimensional hyperbolic metamaterial at $\omega=0$.
 (c): The isofrequency surface of the three-dimensional hyperbolic metamaterial with $\varepsilon_{zz}=\varepsilon_{xx}=\varepsilon_{yy}=1$, and $\mu_{xx}=\mu_{yy}=\mu_{zz}=1$.
These figures are adapted with permission from Ref.~\cite{IYH_PRB(2021)}. Copyright 2024 American Physical Society.
}
 \label{fig:HMM3}
\end{figure}

The above analysis can be extended to a three-dimensional hyperbolic metamaterial.
This hyperbolic metamaterial is described by the $6\times 6$ Maxwell equation,
\begin{equation}
\label{Eq:3d HMM}
\begin{pmatrix}
0&\bm{k}\times\\
-\bm{k}\times&0
\end{pmatrix}
\begin{pmatrix}
\bm{E}\\
\bm{H}
\end{pmatrix}
=\omega
\begin{pmatrix}
\varepsilon&0\\
0&\mu
\end{pmatrix}
\begin{pmatrix}
\bm{E}\\
\bm{H}
\end{pmatrix},
\end{equation}
with
\begin{equation}
\label{eq:3d HMM ep mu}
\varepsilon=
\begin{pmatrix}
1&0&0\\
0&1&0\\
0&0&-1
\end{pmatrix}, \  \  \  \ \mathrm{and} \  \  \  \
\mu = 
\begin{pmatrix}
1&0&0\\
0&1&0\\
0&0&1
\end{pmatrix}.
\end{equation}
We choose the $z$-direction as the anisotropic axis. The eigenvalues are given by
\begin{equation}
\label{eq:EV 3d HMM}
\omega = \pm\sqrt{k_x^2+k_y^2-k_z^2}.
\end{equation}
As is the case with the two-dimensional system,
zero modes are discarded because they do not satisfy Gauss's law. 
The isofrequency surface of the band structure is plotted in Fig.~\ref{fig:HMM3}(a). 
It forms a hyperboloid in the three-dimensional parameter space. 
In this system, an SPES appears in $\omega=0$ [see Fig.~\ref{fig:HMM3}(b)], forming a cone structure. 
This figure indicates that the SPESs are characterized by this zeroth Chern number.
Therefore, SPESs are topologically protected and robust. 
In the three-dimensional model, SPESs vanish when $\varepsilon_{zz}$ becomes positive. 
Corresponding to the vanishing SPESs, the hyperbolic dispersion also vanishes. Figure~\ref{fig:HMM3}(c) shows the isofrequency surface for positive $\varepsilon_{zz}$, elucidating that the isofrequency surface forms an ellipsoid in the momentum space.

In the above, we have demonstrated the emergence of SPERs and SPESs arising from the indefinite property of GEVEs. 
These SPERs or SPESs provide the topological understanding of the hyperbolic dispersion observed in the isofrequency surfaces of hyperbolic metamaterials. 
In addition, we elucidated that the hyperbolic dispersion of hyperbolic metamaterials is related to SPERs and SPESs.
Here, we note that the frequency dependence of permittivity ($\varepsilon$) and permeability ($\mu$) prevents us from the direct observation.
In the following section, we propose an alternative optical system to observe SPERs.

\subsection{SPERs in photonic crystals}\label{sec:GEVE5}

\begin{figure*}[t]
   \includegraphics[width=18cm]{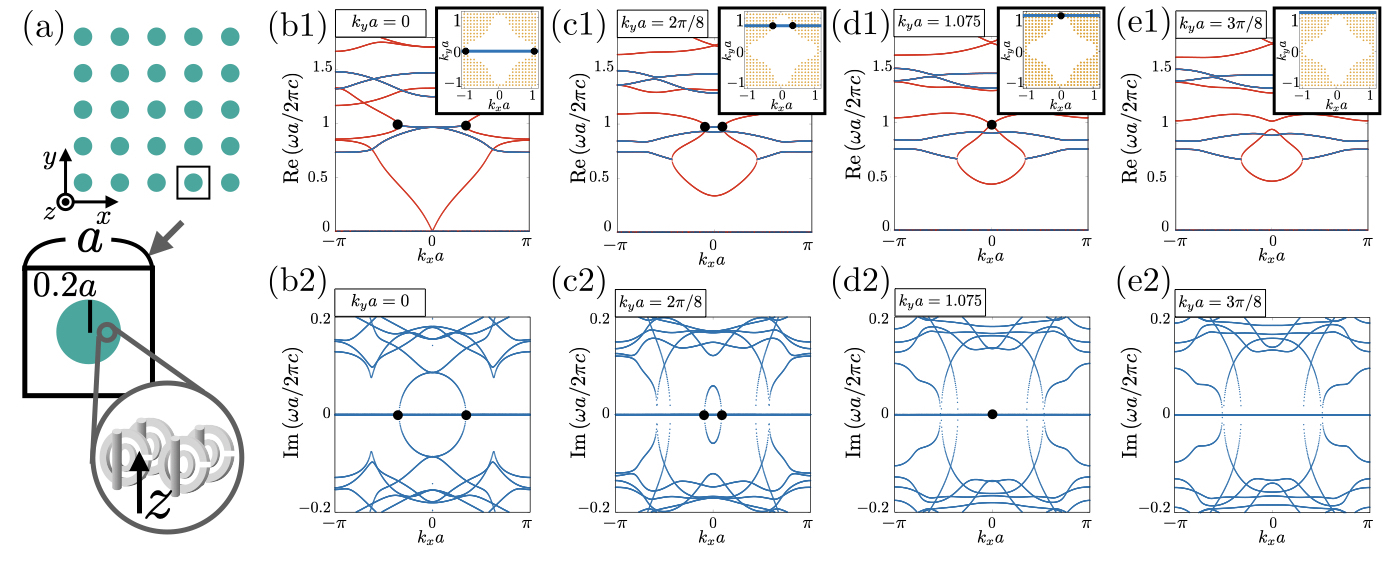}
 \caption{
   (a): Sketch of the photonic crystal composed of negative index media.
   The radius of the internal structures is fixed at $0.2a$.
   (b1)-(e1) [(b2)-(e2)]: Plot of the real [imaginary] part of band structures for $k_ya=0$, $k_ya=2\pi/8$, $k_ya=1.075$, and $k_ya=3\pi/8$ respectively.
Real eigenvalues are plotted in red.
Black dots represent EPs.
The zeroth Chern number is plotted in the insets.
These figures are adapted with permission from Ref.~\cite{IYH_NanoPh(2023)}. Copyright 2024 Walter de Gruyter.

}
 \label{fig:NIM-PhC}
\end{figure*}

\begin{figure}[t]
   \includegraphics[width=7.9cm]{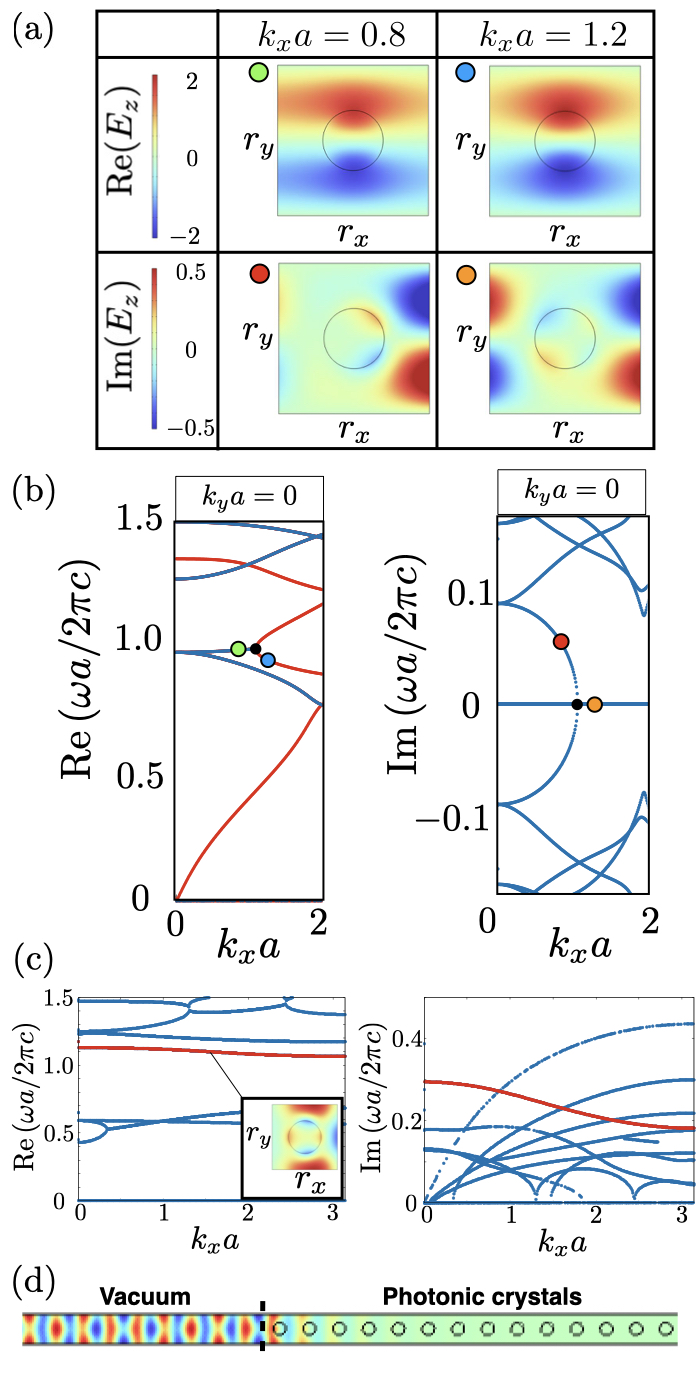}
 \caption{
   (a): Color maps of the real part and the imaginary part of $E_z$ inside and outside of SPER near $\mathrm{Re}[\omega a/2\pi c] = 1$ with $R=0.2a$. 
   (b): Eigenvalues corresponding to the eigenmodes shown in panel (a).
   (c): Band structure for $R = 0.25a$.
   (d): Frequency domain analysis of the photonic crystal composed of negative index media. In order to excite the eigenmode associated with complex eigenvalues using a plane wave, we focus on the complex bands emerging near $\mathrm{Re}[\omega a/2\pi c] \approx 1.1$ for $R = 0.25a$ [red band in panel (c)]. Since the eigenmode is symmetric, it can be excited by the plane wave.
}
 \label{fig:Ez}
\end{figure}

We consider a photonic crystal, an artificial system fabricated by periodically arranged insulators or other materials~\cite{John_PRL(1987),Yablonovitch_PRL(1987),Joannopoulos_APL(1992),Joannopoulos_PhC(1997)}. 
This type of system is one of the most prosperous platforms of the topological photonics as electromagnetic fields can be highly controlled in photonic crystals.

We focus specifically on negative index media where both $\varepsilon$ and $\mu$ take negative values~\cite{Veselago_NIM(2003)}, leading the indefinite matrices of GEVEs.
Such negative index media were first experimentally realized employing thin wire structures and split ring resonators~\cite{Smith_NIM_PRL(2000),Smith_NIM_Science(2001),Smith_NIMreview_Science(2004)} where $\varepsilon$ and $\mu$ are negative for the microwave frequency.
Negative index media of visible lights is realized by employing a fishnet structure~\cite{Dolling_Fishnet_OpticsLetter(2005),Shalaev_Fishnet_Optica(2005),Gunnar_Fishnet_science(2006),Dolling_Fishnet_Optica(2007),Chettiar_Fishnet_OptLetter(2007),Garcia-Meca_Fishnet_OptLett(2009),Xiao_Fishnet_OptLett(2009),Garcia-Meca_Fishnet_PRL(2011),Soukoulis_NIM_NatPhoto(2011)}~\cite{fishnet_ftnt} or by employing solely dielectrics.~\cite{Stephen_D-NIM_JOP(2002),Holloway_D-NIM_IEEE(2003),Peng_D-NIM_PRL(2007),Schuller_D-NIM_PRL(2007),Ahmadi_D-NIM_PRB(2008),Xiaobing_D-NIM_Metamaterials(2008),Vynck_D-NIM_PRL(2009),Soukoulis_NIM_NatPhoto(2011)}~\cite{diele_ftnt}

Let us analyze the transverse magnetic (TM) modes in photonic crystals composed of negative index media [see Fig.~\ref{fig:NIM-PhC}(a)].
For negative index media, we use the composite metamaterial of split-ring resonators and metal wire structures discussed in Ref.~[\citen{Smith_NIM_Science(2001)}]. 
The response of negative index media to the electromagnetic field is incorporated into $\varepsilon$ and $\mu$ using the long-wavelength approximation.
The band structure of the photonic crystal can be obtained by solving the following generalized eigenvalue equation,
\begin{equation}
\label{eq:maxwell eq for phc1}
\sum_j\langle\phi_i|\bm{\nabla}\times\mu^{-1}(\omega_c)\bm{\nabla}\times|\phi_j\rangle\psi_j=\sum_j\left(\frac{\omega a}{2 \pi c}\right)^2\langle\phi_i|\varepsilon(\omega_c)|\phi_j\rangle\psi_j,
\end{equation}
where $\psi_j$ is the eigenvector.
Subscript $j$ specifies a set $(\bm{r}_{\mathrm{c}},\bm{r})$ with $\bm{r}_{\mathrm{c}}$ denoting the position of a cylinder and $\bm{r}=(r_x,r_y)$ denoting the position inside the unit cell. The latter is discritized into linear triangular elements.
Given that $\omega$ is fixed at $\omega_c$, Eq.~\eqref{eq:maxwell eq for phc1} is an eigenvalue problem for the lattice constant $a$. 
We set $\varepsilon(\omega_c)=-5.9$ and $\mu(\omega_c)=-0.4$ by assuming $\omega_c/2\pi=10.7[\mathrm{GHz}]$, indicating that both matrices in Eq.~\eqref{eq:maxwell eq for phc1} are indefinite.

Here, we consider a photonic crystal with a square lattice structure. The radius of the internal structure composed of negative index media is set to $R = 0.2a$ [see Fig.~\ref{fig:NIM-PhC}(a)].
Figure~\ref{fig:NIM-PhC} displays the photonic band structures for the TM mode, characterized by $\bm{E} = (0, 0, E_z)$ and $\bm{H} = (H_x, H_y, 0)$. 
Eigenvalues are computed for each $k_xa$ with the assumption that $\epsilon$ and $\mu$ are constants. 
In Figs.~\ref{fig:NIM-PhC}(b1)-\ref{fig:NIM-PhC}(e1) [\ref{fig:NIM-PhC}(b2)-\ref{fig:NIM-PhC}(e2)], the real (imaginary) parts of the dimensionless parameter $\omega_c a / 2\pi c$ are plotted for different values of $k_y a$, as shown in the insets. 
Bands of real eigenvalues are shown in red. 
Band touching, marked by black dots, is observed in both the real and imaginary parts, indicating the presence of EPs at fixed $k_y a$, as seen in Figs.~\ref{fig:NIM-PhC}(b1)-\ref{fig:NIM-PhC}(d1) and \ref{fig:NIM-PhC}(b2)-\ref{fig:NIM-PhC}(d2). These results suggest the existence of SPERs in a two-dimensional parameter space~\cite{footnote4} [see insets of Figs.~\ref{fig:NIM-PhC}(b1)-\ref{fig:NIM-PhC}(d1)]. 
Here, we note that these EPs are absent in Figs.~\ref{fig:NIM-PhC}(e1) and \ref{fig:NIM-PhC}(e2) because the SPER does not intersect the line specified by $k_y a = 3\pi/8$, as shown in the inset of Fig.~\ref{fig:NIM-PhC}(e1).

The SPER is characterized by a $\mathbb{Z}_2$-invariant. 
In order to calculate $\nu$, we select the two bands involved in the SPER. The $\mathbb{Z}_2$-invariant is shown in the inset of Figs.~\ref{fig:NIM-PhC}(b1)-(e1). 
The inset of Fig.~\ref{fig:NIM-PhC}(b1) confirms that the SPER is characterized by the $\mathbb{Z}_2$-invariant, with $\nu$ transitioning from $-1$ to $1$ across the SPER as $k_x a$ increases from $0$ to $\pi$. Based on these findings regarding the band structure and the $\mathbb{Z}_2$-invariant, we conclude that the photonic crystal composed of negative index media hosts the SPER, protected by the emergent symmetry [see Eq.~\eqref{eq:GEVE6}].

Prior to analyzing how complex eigenmodes propagate in the photonic crystal, we discuss symmetry of eigenmodes. 
Figure~\ref{fig:Ez} displays color maps of electric fields against $r_x$ and $r_y$.
In Fig.~\ref{fig:Ez}(a), the real and imaginary parts of $E_z$ are plotted.  
The wave number is fixed at $k_xa=0.8$ and $k_xa=1.2$, which are located inside and outside of the SPER near $\mathrm{Re}[\omega_c a/2\pi c]=1$.
The electromagnetic field distribution corresponds to the colored points on the band structure in Fig.~\ref{fig:Ez}(b).
The real part is symmetric about $r_x=0$  both inside and outside the SPER (green and blue dots). 
However, the imaginary part becomes asymmetric about $r_x=0$ inside the SPER while it remains anti-symmetric about $r_x=0$ outside the SPER (red and orange dots).

Now, we introduce a plane wave into the photonic crystal composed of negative index media to investigate the behavior of complex eigenmodes. Specifically, we consider the complex eigenvalues at $\mathrm{Re}[\omega_c a/2\pi c] \approx 1.1$ in the photonic crystal with the internal structure radius $R=0.25a$.~\cite{R025a_ftnt} 
The eigenvalue of this band colored with red in Fig.~\ref{fig:Ez}(c) is complex across the Brillouin zone, and the eigenmode is symmetric. 
Thus, one might expect that the eigenmodes of the complex eigenvalues are excitable by the plane wave. 
However, Fig.~\ref{fig:Ez}(d) demonstrates that the eigenmode of the complex eigenvalue cannot be physically excited, and that  the plane wave is reflected on the surface of the photonic crystal.

This is because $\omega$ and $a$ are both real; $a$ describing the size of the unit cell cannot be complex.
Figure~\ref{fig:NIM-PhC} illustrates the band structure of the unit cell size $a$, accommodating eigenmodes with real $\omega$ and real $\bm{k}$. It is important to note that in our analysis, both $\omega$ and $\bm{k}$ are fixed as real values. 
In the complex region of the band structure, $a$ would need to be complex to maintain real $\omega$ and real $\bm{k}$, although physically a complex $a$ is not feasible. 
Therefore, the complex region of the band structure signifies areas where the eigenmodes with real $\omega$ and $\bm{k}$ cannot be physically excited. 
This outcome implies that as $\bm{k}$ varies along the band structure, physically excitable eigenmodes cease to exist at specific $\bm{k}$ points. These points correspond to EPs, and they encircle regions devoid of physically excitable eigenmodes with real $\omega$ and real $\bm{k}$. 
Such a lack of bands is a unique characteristic of the generalized eigenvalue equation involving indefinite Hermitian matrices. 
It is also worth mentioning that spatially decaying modes with complex $\bm{k}$ are not excluded in the complex region. Thus, electromagnetic fields in this context are considered to decay spatially, similar to the behavior observed in photonic band gaps or plasmons in metals.

It is useful to consider experimental observations. In our analysis, the SPER emerges at $\omega_c a/2\pi c \approx 1$. Since our analysis focuses on $\omega_c/2\pi = 10.7\ \text{GHz}$, the unit cell size $a$ of the photonic crystals is determined to be $2.8\ \text{cm}$. 
Thus, one can experimentally access the SPER by preparing a photonic crystal composed of negative index media of this unit cell size.
In the above, we have specifically considered negative index media composed of split-ring resonators and metal-wire structures. However, we consider that the SPER emerges regardless of the composition of the negative index media since we have considered the long-wavelength region.

\begin{figure}[t]
   \includegraphics[width=9cm]{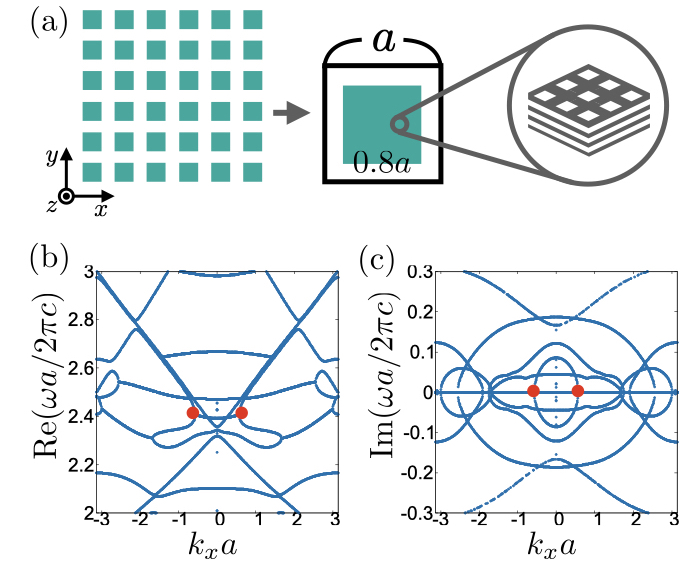}
 \caption{
   (a): Sketch of the photonic crystal composed of Fishnet structure.
   The length of the internal structures is fixed at $0.8a$.
   (b) [(c)]: Plot of the real [imaginary] part of band structures for $k_ya=0$.
We focus on the EPs represented in red dots.
}
 \label{fig:Fishnet}
\end{figure}

Although composite metamaterials have miniaturization limitations, these have been overcome by using fishnet structures. Employing these advanced negative index media might simplify the experimental observation of the SPER. Figure~\ref{fig:Fishnet} shows the result of assuming a fishnet structure as negative index media. Here, we use the values of $\varepsilon$ and $\mu$ obtained from Ref.~[\citen{Garcia-Meca_Fishnet_PRL(2011)}], $\varepsilon = -1$ and $\mu = -1$, neglecting their imaginary parts. The band structure is plotted in Fig.~\ref{fig:Fishnet}(b) and \ref{fig:Fishnet}(c). We focus on the EPs plotted in red, which emerge in the $\omega a/2\pi c \approx 2.4$. Since $\omega/2\pi c = 1/750\ \text{nm}$, the unit cell size is determined to be $1800\ \text{nm}$. This setup is more feasible than the one for composite metamaterials.

\section{Topological photonics in NLEVEs}\label{sec:NLEVP0}

For photonic systems, permittivity and permeability can be frequency dependent [see Eq.~\eqref{eq:maxwell eq for phc}].
For example, this frequency dependence cannot be neglected in photonic systems containing metals~\cite{Sakoda_metal-PhCI_PRB(2001),Sakoda_metal-PhCII_PRB(2001)}. 
Such photonic systems with frequency-dependent $\varepsilon$ and $\mu$ are mathematically described by nonlinear eigenvalue equations (NLEVEs),
\begin{equation}
\label{Eq:NLGEVE}
H(\omega,\bm{k})\bm{\psi}=\omega S(\omega,\bm{k})\bm{\psi},
\end{equation}
which are 
nonlinear with respect to the eigenvalue.~\cite{footnote6}

Chiral edge modes are reported for photonic systems composed of dispersive media.
However, as the frequency dependence is neglected for topological characterization, the robustness of these edge modes remains unclear. 
In this section, introducing auxiliary bands, we elucidate the nonlinear bulk-edge correspondence and clarify the robustness of chiral edge modes under ``weak" nonlinearity.

\subsection{NLEVE and Auxiliary eigenvalue}\label{sec:NLEVP2}
In order to discuss the bulk-edge correspondence in NLEVEs, we introduce the auxiliary eigenvalues.
The use of auxiliary eigenvalues is highly effective in exploring the bulk-edge correspondence in NLEVEs. 
Our strategy provides a way to discuss the nonlinear bulk-edge correspondence under weak but finite nonlinearity.

Here, let us consider the systems described by Eq.~\eqref{Eq:NLGEVE}. 
The topology of such systems can be analyzed by introducing auxiliary eigenvalues $\lambda$, eigenvalues of the matrix $P$ defined as~\cite{IYH_NLEVP_PRL(2024),YIH_arxiv(2024)},
\begin{equation}
\label{Eq:matrix p}
P(\omega,\bm{k})=H(\omega,\bm{k})-\omega S(\omega,\bm{k}).
\end{equation}
The eigenvalues $\omega$ and eigenstates $\bm{\psi}(\omega,\bm{k})$ are obtained as the solution satisfying $\lambda=0$ with
\begin{equation}
\label{eq:aux EV}
P(\omega,\bm{k})\bm{\psi}=\lambda\bm{\psi}.
\end{equation}

When the parameter space described by $\bm{k}$ is $N$-dimensional, we can derive auxiliary band structures in an $N+1$-dimensional parameter space.
Our approach first involves examining the band structure of $\lambda$. 
Then, focusing on the data where $\lambda=0$, we extract the physical band structure. 
The final step is to establish the connection between the auxiliary band structure and the physical band structure.

\begin{figure}[t]   
\includegraphics[width=9cm]{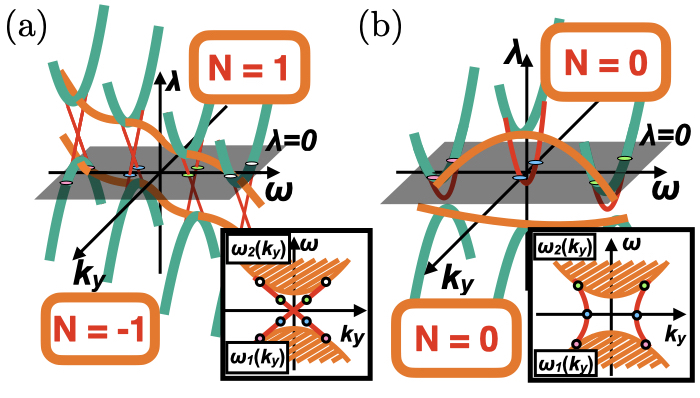}
 \caption{
(a): Sketch of band structures when $\lambda$ is monotonic for each $\omega$.
Bulk (boundary) states are illustrated in green (red).
The band indices of $\lambda$ correspond to the band indices of $\omega$.
The band structure of $\omega$ is sketched in insets.
(b): Band structures when $\lambda$ is not monotonic for each $\omega$.
The band indices of $\lambda$ do not correspond to the band indices of $\omega$.
Although the edge states of $\lambda$ are gapped, edge states of $\omega$ are gapless.
These gapless boundary states cannot be characterized by our method.
These figures are adapted with permission from Ref.~\cite{IYH_NLEVP_PRL(2024)}. Copyright 2024 American Physical Society.
}
\label{fig:AuxBBC}
\end{figure}

We begin with the equation for the auxiliary eigenvalue $\lambda$ [see Eq.~\eqref{eq:aux EV}].
For a two-dimensional system, $\bm{k}$ is also two-dimensional. 
Our focus is on the band gap that includes $\lambda=0$.
Initially, we consider scenarios where the nonlinearity of $\omega$ is weak (i.e., the auxiliary eigenvalues $\lambda$ change monotonically with $\omega$). 
When a topological number calculated from the auxiliary eigenvector is non-zero, gapless boundary states appear around the spatial boundary (e.g., the boundary in the $x$-direction) within the $\lambda$ band gap.
Figure~\ref{fig:AuxBBC}(a) illustrates this occasion.
The bulk band is illustrated in green. 
The red lines in Fig.~\ref{fig:AuxBBC}(a) represent the topological boundary states in the $\lambda$-$k_y$ space for various $\omega$ values. 
Importantly, since these topological boundary states are gapless, they inevitably intersect with $\lambda=0$. This intersection signifies that the topological boundary states of $\lambda$ also manifest as physical boundary states.
Based on the above discussion and the monotonicity of $\lambda$ for each $\omega$, the gapless boundary states are inherited from the auxiliary band structure to the physical one [see inset of Fig.~\ref{fig:AuxBBC}(a)].
Therefore, the bulk-boundary correspondence emerges between the topological number calculated from the auxiliary eigenstates and the physical boundary states.

Here, let us consider the situation involving strong nonlinearity, where the auxiliary eigenvalues $\lambda$ change nonmonotonically with $\omega$.
In general, this strong nonlinearity causes complex $\lambda$. 
Moreover, the strong nonlinearity causes the gapless physical edge states from the topologically trivial auxiliary band structures [see Fig.~\ref{fig:AuxBBC}(b)].
Since the physical gapless edge states cannot be characterized using our method, we discuss the case of weak nonlinearity in the following.

\subsection{
Nonlinear bulk-edge correspondence in a Chern insulator
}
\label{sec:NLEVP3}

\begin{figure*}[h]   
\includegraphics[width=18cm]{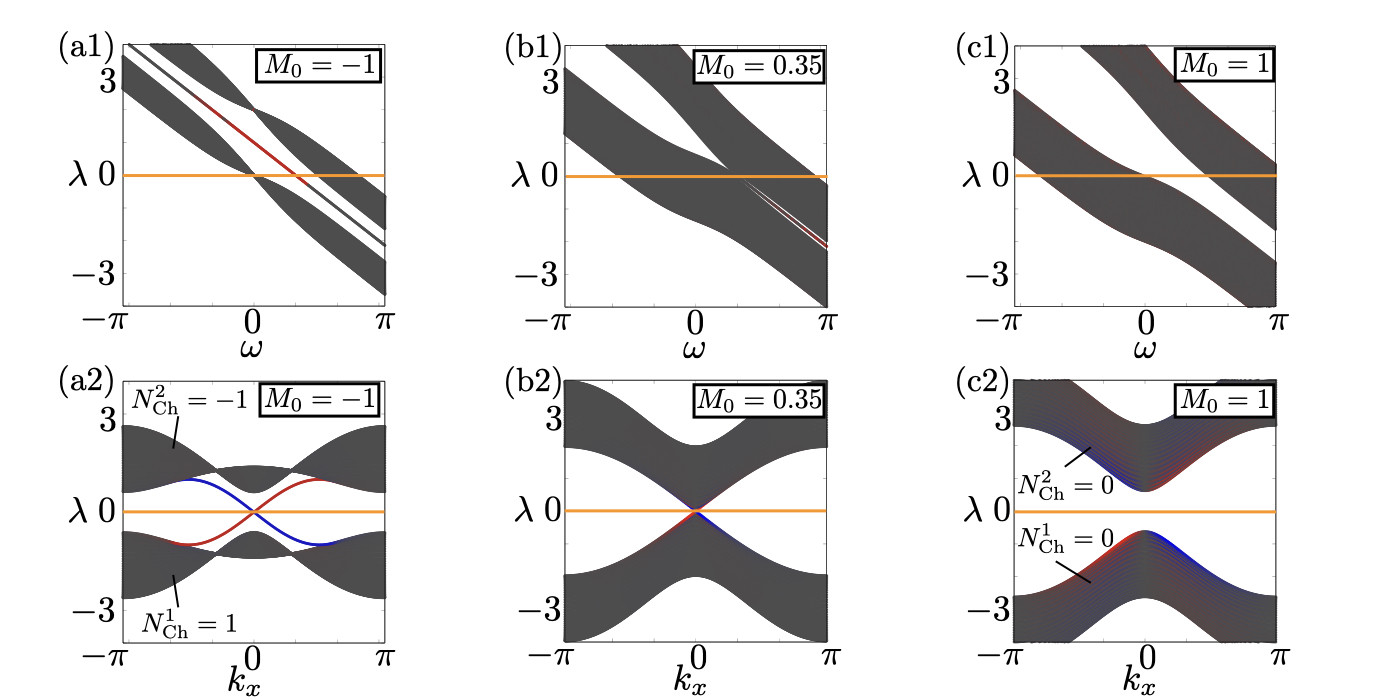}
 \caption{
(a1)-(c1): Auxiliary band structures of $\lambda$ for each $\omega$ at $M_0=-1$, $M_0=0.35$, and $M_0=1$ respectively.
Bulk (boundary) states are plotted in gray (red and blue).
The yellow lines represent $\lambda=0$.
(a2)-(c2): Auxiliary band structures of $\lambda$ for each $k_x$ at $M_0=-1$, $M_0=0.35$, and $M_0=1$ respectively.
These figures are adapted with permission from Ref.~\cite{IYH_NLEVP_PRL(2024)}. Copyright 2024 American Physical Society.

}
\label{fig:NLC1}
\end{figure*}

\begin{figure}[t]   \includegraphics[width=8cm]{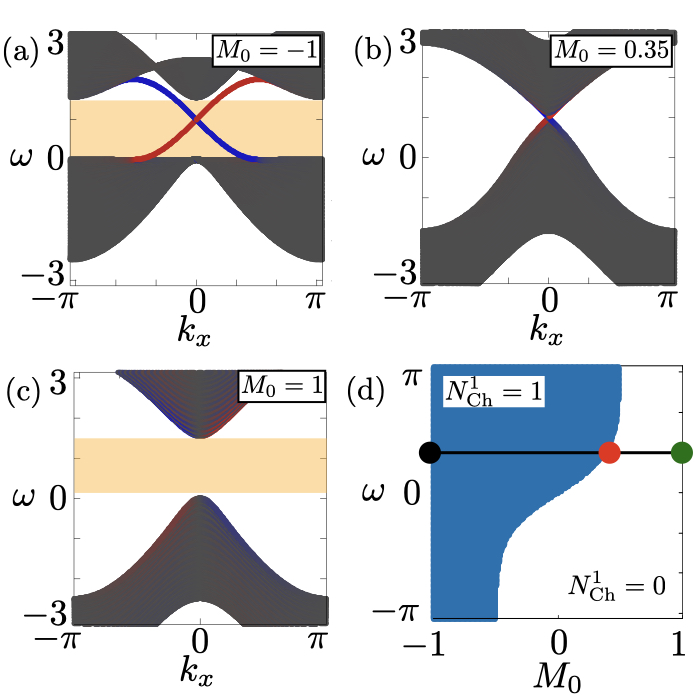}
 \caption{
(a)-(c): Physical band structures of $\omega$ for each $k_x$ at $M_0=-1$, $M_0=0.35$, and $M_0=1$ respectively.
Bulk (boundary) states are plotted in gray (red and blue).
The orange region represents the band gap.
(d): Plot of the Chern number of the lower band.
The blue region and the white region represent areas where the Chern number takes $1$ and $0$.
Black, red, and green dots represent the points where $M_0=-1$, $M_0=0.35$, and $M_0=1$ with $\omega=1$, respectively. 
These figures are adapted with permission from Ref.~\cite{IYH_NLEVP_PRL(2024)}. Copyright 2024 American Physical Society.
}
\label{fig:NLC2}
\end{figure}

In this section, we elucidate that the Chern number of auxiliary bands protects the chiral edge modes.
As an example, we begin with the nonlinear eigenvalue equation given by,
\begin{equation}
\label{Eq:NLEVP in NLCI}
H(\bm{k})\bm{\psi}=\omega S(\omega)\bm{\psi},
\end{equation}
where $H(\bm{k})$ and $S(\omega)$ are matrices depend on both $\bm{k}$ and $\omega$,
\begin{align}
\label{Eq:matrices in NLCI}
H(\bm{k})=
\begin{pmatrix}
E+M_\mathrm{H}(\bm{k})&\sin(k_x)-i\sin(k_y)\\
\sin(k_x)+i\sin(k_y)&E-M_\mathrm{H}(\bm{k})
\end{pmatrix},\\
S(\omega)=
\begin{pmatrix}
1-M_\mathrm{S}(\omega)&0\\
0&1+M_\mathrm{S}(\omega)
\end{pmatrix},
\end{align}
with $E=1$, $M_{\mathrm{H}}(\bm{k})=M_0+\sum_{i=x,y}[1-\cos(k_i)]$, and $M_{\mathrm{S}}(\omega)=M_1\tanh(\omega)/\omega$. Here, $M_1$ is fixed to $-0.5$.

We define the matrix $P$ as 
\begin{equation}
\label{Eq:P in NLCI}
P(\omega,\bm{k})=
\begin{pmatrix}
E_{\mathrm{P}}(\omega)+M_{\mathrm{P}}(\omega,\bm{k})&\sin(k_x)-i\sin(k_y)\\
\sin(k_x)+i\sin(k_y)&E_{\mathrm{P}}(\omega)-M_{\mathrm{P}}(\omega,\bm{k})
\end{pmatrix},
\end{equation}
with 
$E_{\mathrm{P}}(\omega)=E-\omega$ and $M_{\mathrm{P}}(\omega,\bm{k})=M_{\mathrm{H}}(\bm{k})+\omega M_{\mathrm{S}}(\omega)$.
We investigate the auxiliary band structure by solving $P(\omega,\bm{k})\bm{\psi}=\lambda\bm{\psi}$. The resulting band structure is illustrated in Fig.~\ref{fig:NLC1}. 
The data are obtained under open boundary conditions along the $y$-axis and periodic boundary conditions along the $x$-axis. Figures~\ref{fig:NLC1}(a1) through \ref{fig:NLC1}(c1) [\ref{fig:NLC1}(a2) through \ref{fig:NLC1}(c2)] depict the $\lambda$ bands as functions of $\omega$ [$k_x$] with $k_x=0$ [$\omega=1$]. The bulk states are colored with gray, while the boundary states are marked in red.

For $M_0=-1$, boundary states appear as seen in Figs.~\ref{fig:NLC1}(a1) and \ref{fig:NLC1}(a2). These boundary states are gapless, crossing $\lambda=0$, which indicates the presence of physical boundary states inherited from the auxiliary bands. In Figs.~\ref{fig:NLC1}(b1) and \ref{fig:NLC1}(b2), the band gap closes near $M_0=0.35$, forming a Dirac point at $\lambda(\omega=1,k_x=0)$. As $M_0$ increases to $1$, the gap reopens and the boundary states vanish, signaling a topological phase transition near $M_0 = 0.35$. Thus, for $M_0 < 0.35$, physical boundary states are expected due to the inevitable crossing of $\lambda=0$ by the auxiliary boundary states.

Examining topology of the band structure of $\lambda(\omega,k_x)$ elucidates the robustness of the gapless boundary modes $\omega(k_x)$ [Fig.~\ref{fig:NLC1}].
The Chern number, serving as this topological invariant, is defined as~\cite{Chernnumber_ftnt}
\begin{equation}
\label{Eq:NLChern num}
N_{\mathrm{Ch}}^{n}(\omega) = \frac{1}{2\pi} \int_{\mathrm{1BZ}} dk_x dk_y \bm{\nabla}_k \times \bm{A}_n(\omega, \bm{k}),
\end{equation}
\begin{equation}
\label{Eq:NLBerry}
\bm{A}_n(\omega, \bm{k}) = \langle \psi_{n,k}(\omega) | \bm{\nabla}_k \psi_{n,k}(\omega) \rangle,
\end{equation}
where $n$ denotes the band index.
The integration is over the first Brillouin zone in momentum space. The Berry connection $\bm{A}_n(\omega, \bm{k})$ depends on $\omega$, making the Chern number a function of $\omega$. It is crucial to note that the Chern number is calculated from the eigenstates of $\lambda$, distinguishing it from the Chern number obtained via self-consistent analysis.

In our model, a non-zero Chern number is observed for $M_0<0.35$. In Fig.~\ref{fig:NLC1}(a2), the Chern numbers for the lower and upper bands are $N_{\mathrm{Ch}}^{1}(\omega_{\mathrm{R}}) = 1$ and $N_{\mathrm{Ch}}^{2}(\omega_{\mathrm{R}}) = -1$, respectively, while in Fig.~\ref{fig:NLC1}(c2), both Chern numbers are zero. Thus, the Chern numbers derived from the $\lambda$ eigenstates correspond to the boundary states of the auxiliary bands of $\lambda$.

We then explore the nonlinear bulk-edge correspondence between the Chern number of auxiliary bands and the physical boundary states of $\omega$. The band structures of $\omega$ 
are shown in Fig.~\ref{fig:NLC2}. 
Figures~\ref{fig:NLC2}(a), \ref{fig:NLC2}(b), and \ref{fig:NLC2}(c) display the bands of $\omega(k_x)$ for $M_0=-1$, $M_0=0.35$, and $M_0=1$, respectively. 
The bulk and boundary states are colored gray and red, respectively.  
The region where the $\omega$ band gap remains open for all $k_x$ is shown in orange. 
Gapless boundary states of $\omega$ appear for $M_0=-1$ [see Fig.~\ref{fig:NLC2}(a)], inherited from the auxiliary boundary states of $\lambda$ in Fig.~\ref{fig:NLC2}(a2). 
Near $M_0=0.35$, a gapless point, corresponding to the Dirac point in Fig.~\ref{fig:NLC1}(b2), emerges [see Fig.~\ref{fig:NLC2}(b)]. For $M_0=1$, the boundary states disappear [see Fig.~\ref{fig:NLC2}(c)].

Figure~\ref{fig:NLC2}(d) plots the Chern number of the lower band calculated from eigenstates of $\lambda$ for each $M_0$. 
Importantly, there is a correspondence between the regions of non-zero Chern number and the emergence of physical boundary states. 
This indicates the existence of the nonlinear bulk-edge correspondence in our two-dimensional system. 
The reference point $\omega_R$ for the calculation of the Chern number should be selected within the orange-colored region in Figs.~\ref{fig:NLC2}(a)-\ref{fig:NLC2}(c), due to potential band gap closing and topological phase transitions within the white region.
In the analysis of Fig.~\ref{fig:NLC2}(d), the black line represents $\omega_R$ and is chosen to correspond to the frequency of the Dirac point.

The above argument of bulk-edge correspondence for systems described by NLEVEs elucidates the topological protection of chiral edge modes under ``weak" nonlinearity.
We also note that the above analysis of the discussion of the two-dimensional model can be extended to the three-dimensional systems of the Weyl semimetal by replacing $M_0$ with $\mathrm{cos}(k_z)$~\cite{IYH_NLEVP_PRL(2024)}.
In this case, the boundary between the white region and the blue region in Fig.~\ref{fig:NLC2}(d) corresponds to the Weyl points~\cite{Chernnumber2_ftnt}.

We finish this section with comments on the edge modes of photonic crystals.
Applying this argument to photonic systems clarifies the robustness of edge modes.
For instance, chiral edge modes are reported in the photonic crystal made of gyromagnetic materials~\cite{MIT_PhChIns_PRL(2008),Wang_exp_Nat(2009)}.
However, because the $\omega$ dependence of $\varepsilon(\omega)$ and $\mu(\omega)$ was neglected for topological characterization in previous works, the robustness of the chiral edge modes remained unclear (in particular, the robustness against nonlinearity). 
Computing the Chern number of auxiliary bands elucidates the robustness of the chiral edge modes in photonic crystals whose $\varepsilon(\omega)$ and $\mu(\omega)$ are $\omega$ dependent.

\section{Summary}

We have briefly reviewed the topological photonics in the context of GEVEs and NLEVEs. 
Photonic systems are precisely described by GEVEs or NLEVEs, which are beyond conventional topological band theory.

First, we have discussed the complex band structure induced by the indefinite property of matrices in GEVEs. 
These complex bands preserve the emergent symmetry that arises from the GEVE and the Hermiticity of the matrices.
This symmetry allows EPs to emerge as higher-dimensional structures,
which explains the characteristic dispersion relation of the hyperbolic metamaterial.
Moreover, we have applied these discussions to photonic systems such as hyperbolic metamaterials and photonic crystals composed of negative index media. 

For systems of NLEVEs, we have shown that the bulk-edge correspondence holds under weak but finite nonlinearity of eigenvalues by introducing auxiliary eigenvalues. 
Although the auxiliary bands lose their physical meaning away from $\lambda=0$, their topology clarifies the robustness of the physical edge modes, which is applicable to photonic systems with dispersive materials.
We note that the auxiliary eigenvalues not only provide insights into NLEVEs but also offer a new perspective on the discussion of GEVEs. 
Using auxiliary eigenvalues, the emergence of complex band structures, as discussed in Sec.~\ref{sec:GEVE3}, can be understood through the analysis of conic sections~\cite{IYH_arxiv(2024)}.
Research in topological photonics beyond the standard eigenvalue problem is still in its early stages, and it is a field with great potential for future development.

\begin{acknowledgment}
This work is supported by JST-CREST Grant No.~JPMJCR19T1, JST-SPRING Grant No.~JPMJSP2124, and JSPS KAKENHI Grant No.~JP21K13850, JP23K25788, and JP23KK0247.
This work is also supported by JSPS Bilateral Program No.~JSBP120249925.
T.Y is grateful for the support from the ETH Pauli Center for Theoretical Studies and the Grant from Yamada Science Foundation.
\end{acknowledgment}


\begin{thebibliography}{100}
\bibitem{footnote5}
{ The $\mathbb{Z}_2$ invariant of the discriminant corresponds to the parity of the zeroth Chern number, which can be seen as follows. In the presence of the pseudo-Hermiticity, two eigenvalues are real or are related to each other by complex conjugation. The number of the pairs changes by one at SPERs, leading to the change of the zeroth Chern number by $\pm1$. Because the sign of $\Delta(\bm{k})$ is plus (minus) when the number of the pair is odd (even), the $\mathbb{Z}_2$-invariant $\nu=\mathrm{sgn}(\Delta)$ corresponds to the parity of the zeroth Chern number }.
\bibitem{R025a_ftnt}
{ In the two-dimensional simulation with $R=0.2a$, even when a plane wave with $\omega a/2\pi c=1$ is incident on the SPER, there exist surrounding real eigenvalues in a ring-like region, which are unintentionally excited along with the eigenmodes of the complex eigenvalues. To avoid this situation and focus solely on the eigenmodes associated with complex eigenvalues within the SPER, we perform simulations using a structure with $R=0.25a$, where the complex eigenvalues extend over the entire first Brillouin zone, corresponding to a SPER whose diameter exceeds that of the Brillouin zone }.
\bibitem{footnote6}
{ The definite property of the matrices in Eq.~\eqref{Eq:NLGEVE} depends on the value of $\omega$, making the discussion of generalized eigenvalue problems from the previous sections inapplicable. Therefore, in this section, we discuss the NLEVE using a different approach }.
\bibitem{Chernnumber_ftnt}
{ Here, the Chern number is computed for ``occupied states" of $\lambda$-bands. The Chern number predicts physical edge states when $\lambda=0$ exist within the band gap of $\lambda$ }.
\bibitem{Chernnumber2_ftnt}
{ We note that this correspondence depends on the details of the model. For instance, in general, the Chern number may change via gap-closing at finite $\lambda$. In this case, $\omega$-bands become metallic rather than semi-metallic }.
\end{thebibliography}
\end{document}